%% file: main.tex
\documentclass{vldb}

\usepackage[caption=false]{subfig}  % font=normalsize
\usepackage{graphicx}
\usepackage[usenames]{color}

\usepackage{multicol}
\usepackage{syntax}
\usepackage{booktabs}
\usepackage[normalem]{ulem}

\usepackage{latexsym}
\usepackage{stmaryrd}

\usepackage{pstricks}
\usepackage{pst-node}
\usepackage{pst-tree}

\usepackage{paralist}
\usepackage{url}

\usepackage{balance}

\usepackage{algorithmic}

\newtheorem{theorem}{Theorem}

\newtheorem{proposition}{Proposition}

\newdef{definition}{Definition} 
\newdef{example}{Example} 
\newdef{algorithm}{Algorithm}

\newcommand{\T}{\mathcal{T}}
\newcommand{\U}{\mathcal{U}}
\newcommand{\V}{\mathcal{V}}
\newcommand{\A}{\mathcal{A}}
\newcommand{\B}{\mathcal{B}}

\newcommand{\D}{\mathbf{D}}

\renewcommand{\phi}{\varphi}
\newcommand{\tuple}[1]{{\langle#1\rangle}}
\newcommand{\dbracket}[1]{{\llbracket#1\rrbracket}}

\newcommand{\bttuple}[1]{{\langle\textcolor{blue}{\mathrm{#1}}\rangle}}
\newcommand{\data}[2]{\langle\textcolor{blue}{{#1}\!:\!{#2}}\rangle}

\newcommand{\agg}[1]{\gamma_{#1}}

\def\punto{$\hspace*{\fill}\Box$}

\newcommand{\new}[1]{{\color{magenta}{#1}}}

\newcommand{\nop}[1]{}

\begin{document}

\title{Aggregation and Ordering in Factorised Databases}

\numberofauthors{1}

\author{
\alignauthor
Nurzhan Bakibayev, Tom\'{a}\v{s} Ko\v{c}isk\'{y}, Dan Olteanu, and Jakub Z\'{a}vodn\'{y}\\
       \affaddr{Department of Computer Science, University of Oxford, OX1 3QD, UK}
       \email{\{nurzhan.bakibayev, tomas.kocisky, dan.olteanu, jakub.zavodny\}@cs.ox.ac.uk}
}

\maketitle
\begin{abstract}
  A common approach to data analysis involves understanding and
  manipulating succinct representations of data. In earlier work, we
  put forward a succinct representation system for relational data
  called factorised databases and reported on the main-memory query
  engine FDB for select-project-join queries on such databases.

  In this paper, we extend FDB to support a larger class of practical
  queries with aggregates and ordering. This requires novel
  optimisation and evaluation techniques. We show how factorisation
  coupled with partial aggregation can effectively reduce the number
  of operations needed for query evaluation. We also show how
  factorisations of query results can support enumeration of tuples in
  desired orders as efficiently as listing them from the unfactorised,
  sorted results.

  We experimentally observe that FDB can outperform off-the-shelf
  relational engines by orders of magnitude.
\end{abstract}

\nop{
\category{H.2.4}{Information Systems Applications}{Miscellaneous}
%A category including the fourth, optional field follows...
 \category{D.2.8}{Software Engineering}{Metrics}[complexity measures, performance measures]

\terms{Theory}

\keywords{}
}

%%%%%%%%%%
\input{introduction}
\input{prelims}

\input{aggregation}

\input{gby-oby}

\input{optimisation}
\input{experiments}

\input{relatedwork}

\input{conclusion}

%%%%%%%%%%

\balance

\bibliographystyle{abbrv}
%{\small
\bibliography{bibtex}
%}

\balance

\end{document}

%% file: introduction.tex
% !TEX root = main.tex

\section{Introduction}

Succinct representations of data have been developed in various
fields, including computer science, statistics, applied mathematics,
and signal processing. Such representations are employed among others
for capturing data during measurements, as used in compressed sensing
and sampling, and for storing and transmitting otherwise large amounts
of data, as used in signal analysis, statistical analysis, complex
query processing, and machine learning and
optimisation~\cite{workshop:datarepr:2013}. They can speed up data
analysis and in some cases even bring large-scale tasks into the realm
of the feasible.

In this paper, we consider the evaluation problem for queries with
aggregates and ordering on a succinct representation of relational
data called {\em factorised databases}~\cite{OZ12}. 

This representation system uses the distributivity of product over
union to factorise relations, similar in spirit to factorisation of
logic functions~\cite{brayton87}, and to boost the performance of
relational processing~\cite{BOZ12}. It naturally captures as a special
case lossless decompositions defined by join dependencies, as
investigated in the context of normal forms in database
design~\cite{AHV95}, conditional independence in Bayesian
networks~\cite{Pearl:Prob:1989}, minimal constraint networks in
constraint satisfaction~\cite{Gottlob11}, and in our previous work on
succinct representation of query results~\cite{OZ12} and their
provenance polynomials~\cite{OZ11b} used for efficient computation in
probabilistic databases~\cite{OH2008,Prithvi:VLDB:2010}. It also
captures product decompositions of relations as studied in the context
of incomplete information~\cite{OKA08gWSD}, as well as factorisations
of relational data representing large, sparse feature matrices
recently used to scale up machine learning
algorithms~\cite{Rendle:PVLDB:2013}. These existing decomposition
techniques can be straightforwardly used to supply data in factorised
form.

In earlier work, we introduced the FDB main-memory engine for
select-project-join queries on factorised databases~\cite{BOZ12}
and showed that it can outperform relational engines by orders of
magnitude on data sets with many-to-many relationships. In this paper,
we extend FDB to support a larger class of practical queries with
(sum, count, avg, min, max) aggregates, group-by and order-by clauses,
while still maintaining its performance superiority over relational
query techniques.

Factorisation can benefit aggregate computation. For instance,
counting tuples of a relation factorised as a union of products of
relations can be expressed as a sum of multiplications of
cardinalities of those latter relations. The reduction in the number
of computation steps brought by factorisation over relational
representation follows closely the gap in the representation size and
can be arbitrarily large; we experimentally show performance
improvements of orders of magnitude. Further speedup is achieved by
evaluating aggregation functions as sequences of repeated partial
aggregations on factorised data, possibly intertwined with
restructuring of the factorisation.

For queries with order-by clauses, there are factorisations of query
results for which their tuples can be enumerated in desired orders
with the same time complexity (constant per tuple) as listing them
from the sorted query results. Any factorisation can be restructured
so as to support constant-delay enumeration in a given order. This
restructuring is in most cases partial and builds on the intuition
that, even in the relational case, sorting can partially use an
existing order instead of starting from scratch. For instance, if a
relation is sorted by A,B,C, re-sorting by B,A,C need not re-sort the
C-values for any pair of values for A and B.

%%%%%%%%%%%%%%%%%%%%
\begin{figure*}[t]
\begin{center}
\begin{scriptsize}
\subfloat{
\begin{tabular}{@{}l@{~}@{~}l@{~}@{~}l@{~}@{~}}
\multicolumn{3}{c}{\textrm{Orders}}\\\toprule
 customer & date & pizza \\\midrule
 Mario & Monday & Capricciosa \\
 Mario & Tuesday & Margherita \\
 Pietro & Friday & Hawaii \\
 Lucia & Friday & Hawaii \\
 Mario & Friday & Capricciosa \\\bottomrule
\ \\
\ \\
\ \\
\end{tabular}
}\quad
\subfloat{
\begin{tabular}{@{~}l@{~}l@{~}}
\multicolumn{2}{c}{\textrm{Pizzas}}\\\toprule
pizza & item \\\midrule
 Margherita & base \\
 Capricciosa & base \\
 Capricciosa & ham \\
 Capricciosa & mushrooms \\
 Hawaii & base \\
 Hawaii & ham \\
 Hawaii & pineapple \\\bottomrule
\ \\
\end{tabular}
}%
\subfloat{
\begin{tabular}{@{~}l@{~}l@{~}l@{~}}
\multicolumn{2}{c}{\textrm{Items}}\\\toprule
 item & price\\\midrule
 base & 6 \\
 ham & 1 \\
 mushrooms & 1 \\
 pineapple & 2 \\\bottomrule
\ \\
\ \\
\ \\
\ \\
\end{tabular}
}
\subfloat{
\begin{tabular}{l}
 $\bttuple{Capricciosa} \times (\bttuple{Monday} \times \bttuple{Mario} \;\cup$ \\
 $\phantom{\bttuple{Capricciosa}\times(}\bttuple{Friday} \times \bttuple{Mario})$ \\
 $\phantom{\bttuple{Capricciosa}} \times (\bttuple{base}\times \bttuple{6} \;\cup$\\
 $\phantom{\bttuple{Capricciosa} \times (} \bttuple{ham}\times \bttuple{1} \;\cup$ \\
 $\phantom{\bttuple{Capricciosa} \times (} \bttuple{mushrooms}\times\bttuple{1}) \;\cup$\\
 $\bttuple{Hawaii} \times \bttuple{Friday} \times (\bttuple{Lucia} \cup \bttuple{Pietro})$ \\
 $\phantom{\bttuple{Hawaii}} \times (\bttuple{base}\times\bttuple{6} \;\cup$\\
 $\phantom{\bttuple{Hawaii} \times (} \bttuple{ham}\times \bttuple{1} \;\cup$\\
 $\phantom{\bttuple{Hawaii} \times (} \bttuple{pineapple}\times \bttuple{2}) \;\cup$ \\
 $\bttuple{Margherita} \times \bttuple{Tuesday} \times \bttuple{Mario} \times \bttuple{base} \times \bttuple{6}$
\end{tabular}
}
\nop{%
\subfloat{
\begin{tabular}{lllll}
   \multicolumn{5}{c}{$\textrm{Orders}\Join\textrm{Items}\Join\textrm{Pizzas}$}\\\toprule
 customer & date & pizza & item & price \\\midrule
 Mario & Monday & Capricciosa & base & 6 \\
 Mario & Monday & Capricciosa & ham & 1 \\
 Mario & Monday & Capricciosa & mushrooms & 1 \\
 Mario & Tuesday & Margherita & base & 6 \\
 Pietro & Friday & Hawaii & base & 6 \\
 Pietro & Friday & Hawaii & ham & 1 \\
 Pietro & Friday & Hawaii & pineapple & 2 \\
 \multicolumn{5}{c}{$\dots$}\\\bottomrule
\end{tabular}}
}

\end{scriptsize}
\end{center}
\vspace*{-1em}

\caption{From left to right: An example pizzeria database 
(\textrm{Orders}, \textrm{Pizzas}, \textrm{Items}); a factorisation of
the natural join of the three relations, whose nesting structure is
given by the factorisation tree $\T_1$ in Figure~\ref{fig:running-example}. }
\label{fig:ex-db}
\vspace*{-1em}
\end{figure*}
%%%%%%%%%%%%%%%%%%%%

\begin{figure*}[t]
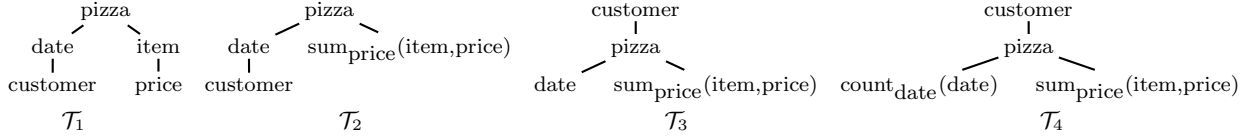

\begin{center}
\begin{small}
\[
\psset{levelsep=5mm, nodesep=2pt, treesep=5mm}
\pstree{\TR{\mbox{pizza}}}
{ \pstree{\TR{\mbox{date}}} { \TR{\mbox{customer}} }
\pstree{\TR{\mbox{item}}} { \TR{\mbox{price}} } }\hspace*{1em}
\pstree{\TR{\mbox{pizza}}}
{
  \pstree{\TR{\mbox{date}}}
  {
    \TR{\mbox{customer}}
  }
  \TR{\mbox{sum}_{\mbox{price}}(\mbox{item,price})}
}\hspace*{1em}
\pstree{\TR{\mbox{customer}}}
{
  \pstree{\TR{\mbox{pizza}}}
    {
      \TR{\mbox{date}}
      \TR{\mbox{sum}_{\mbox{price}}(\mbox{item,price})}
    }
}\hspace*{1em}
\pstree{\TR{\mbox{customer}}}
{
  \pstree{\TR{\mbox{pizza}}}
    {
      \TR{\mbox{count}_{\mbox{date}}(\mbox{date})}
      \TR{\mbox{sum}_{\mbox{price}}(\mbox{item,price})}
    }
}
\]
\end{small}\vspace*{-1em}

$\T_1$\hspace*{10em} $\T_2$\hspace*{12em} $\T_3$\hspace*{14em} $\T_4$\hspace*{5em}
\end{center}\vspace*{-1.5em}

\caption{Factorisation trees used Example~\ref{ex:running-example}.}
\label{fig:running-example}
\end{figure*}

\begin{example}\label{ex:running-example}
  Figure~\ref{fig:ex-db} shows a database with pizzas on offer,
  pri\-ces of toppings, and pizza orders by date, as well as a
  factorisation of the relation $R =
  \textrm{Orders}\Join\textrm{Pizzas}\Join\textrm{Items}$. This
  factorisation has the nesting structure $\T_1$
  (Figure~\ref{fig:running-example}) that reflects the join
  dependencies in $R$ as defined by the natural join of the three
  relations.  Such nesting structures are called {\em factorisation
    trees} (f-trees). We read this factorisation as follows. We first
  group by pizzas; for each pizza, we represent the orders separately
  from toppings and prices.

  We next present three scenarios of increasing complexity where
  factorisation can benefit aggregate computation. We assume the
  factorisation of the materialised view $R$ given. Throughout the
  paper, we express aggregation using the $\varpi_{G;agg}$ operator,
  which groups by attributes $G$ and applies the aggregation function
  $agg$ within each group.

  1. We first consider the case when the aggregation only applies
  locally to a fragment of the factorisation and there is no need to
  restructure the factorisation. An example query would find the price
  of each ordered pizza: $$S = \varpi_{\mbox{customer, date, pizza;
      sum(price)}} (R).$$ We can evaluate this query directly on the
  factorisation of $R$, where for each pizza we replace the
  expressions over items and price by the sum of the prices of all its
  items:\vspace*{-.5em}

\begin{scriptsize}
\hspace*{-1em}\begin{align*}
 &\bttuple{Capricciosa} \times (\bttuple{Monday} \times \bttuple{Mario} \;\cup \bttuple{Friday} \times \bttuple{Mario}) \times \bttuple{8} \;\cup\\
 &\bttuple{Hawaii} \times \bttuple{Friday} \times (\bttuple{Lucia} \cup \bttuple{Pietro}) \times \bttuple{9} \;\cup\\
 &\bttuple{Margherita} \times \bttuple{Tuesday} \times \bttuple{Mario} \times \bttuple{6}
\end{align*}
\end{scriptsize}\vspace*{-1em}

The f-tree of this factorisation is $\T_2$ from
Figure~\ref{fig:running-example}.

2. If the aggregation attributes are distributed over the f-tree, we
may need to restructure the factorisation to be able to aggregate
locally as in the previous example. Alternatively, we can decompose
the aggregation operation into several partial aggregation steps and
intertwine them with restructuring operations. An example query in
this category finds the revenue per customer: $$P =
\varpi_{\mbox{customer; sum(price)}}(R).$$ We first partially
aggregate prices per pizza, see $S$ above.  The factorisation of $S$
is then restructured from the f-tree $\T_2$ to $\T_3$ (see
Figure~\ref{fig:running-example}), so that we first group by
customers: \vspace*{-2em}

\begin{scriptsize}
\begin{align*}
 &\bttuple{Lucia} \times \bttuple{Hawaii} \times \bttuple{Friday} \times \bttuple{9} \cup\\
 &\bttuple{Mario} \times (\bttuple{Capricciosa} \times (\bttuple{Monday} \cup \bttuple{Friday}) \times \bttuple{8} \cup \\
 &\phantom{\bttuple{Mario} \times(} \bttuple{Margherita} \times \bttuple{Tuesday} \times \bttuple{6}) \cup\\
 &\bttuple{Pietro} \times \bttuple{Hawaii} \times \bttuple{Friday} \times \bttuple{9}
\end{align*}
\end{scriptsize}\vspace*{-1em}

For each pizza ordered by a customer, we next count the number of
order dates and obtain the following factorisation over the f-tree
$\T_4$ in Figure~\ref{fig:running-example}:\vspace*{-1em}

\begin{scriptsize}
\begin{align*}
 &\bttuple{Lucia} \times \bttuple{Hawaii} \times \bttuple{1} \times \bttuple{9} \cup\\
 &\bttuple{Mario} \times (\bttuple{Capricciosa} \times \bttuple{2} \times \bttuple{8} \cup \\
 &\phantom{\bttuple{Mario} \times(} \bttuple{Margherita} \times \bttuple{1} \times \bttuple{6}) \cup\\
 &\bttuple{Pietro} \times \bttuple{Hawaii} \times \bttuple{1} \times \bttuple{9}
\end{align*}
\end{scriptsize}\vspace*{-1em}

This is a further example of partial aggregation that helps prevent
possibly large factorisations. Finally, we compute the revenue per
customer by aggregating the whole subtree under customer. This is
accomplished by first computing the revenue per pizza, which is
obtained by multiplying the partial count and sum
aggregates\nop{(f-tree not shown)} and then summing over all pizzas
for each customer. The final result is:\vspace*{-1em}

\begin{scriptsize}
\begin{align*}
 &\bttuple{Lucia} \times \bttuple{9} \cup \bttuple{Mario} \times \bttuple{22} \cup \bttuple{Pietro} \times \bttuple{9}
\end{align*}
\end{scriptsize}\vspace*{-1.2em}

The f-tree of this factorisation is: \vspace*{-0.8em}

\begin{small}
\[
\hspace*{1em}
\psset{levelsep=5mm, nodesep=2pt, treesep=5mm}
\pstree{\TR{\mbox{customer}}}
{ \TR{\mbox{sum}_{\mbox{price}}(\mbox{item,price,pizza,date})} }
\]
\end{small}\vspace*{-.5em}

3. If we are interested in enumerating the tuples in query results
with constant delay, as opposed to materialising their factorisations,
we can avoid several restructuring steps and thus save computation.
For instance, if we would like to compute the revenue per customer and
pizza, we could readily use the factorisation over $\T_4$, since for
each customer we could multiply the partial aggregates for the number
of order dates and the price per pizza for each pair of customer and
pizza on the fly. In general, if all group-by attributes are above the
other attributes in the f-tree of a factorised relation, then we can
enumerate its tuples while executing partial aggregates on the other
attributes on the fly.\punto
\end{example}

For queries with order-by clauses, the tuples in the factorised query
result can be enumerated with constant delay in the order expressed in
the query if two conditions are satisfied: (i) as for group-by
clauses, all order-by attributes are above the other attributes in the
f-tree of the factorisation, and (ii) the sorting order expressed by
the order-by clause is included in some topological order of the
f-tree.

\begin{example}
  Factorisations over $\T_1$ can support the orders: (pizza); (pizza,
  date); (pizza, item); (pizza, item, date); (pizza, date, item); and
  so on. The order (customer, pizza, item, price) can be obtained by
  pushing up the customer attribute past attributes date and pizza in
  the f-tree $\T_1$; this, however, need not change the factorisation
  for pizza, items, and price representing the right branch in
  $\T_1$.\punto
\end{example}

Our main observation is that the FDB query engine can clearly
outperform relational techniques if the input data is factorised. This
case fits a read-optimised scenario with views materialised as
factorisations and on which subsequent processing is conducted. The
key performance improvement is brought by the succinctness of
factorisations.

The contribution of this paper lies in addressing the evaluation and
optimisation problems for queries with aggregates and ordering on
factorised databases. In particular:
\begin{compactitem}

\item We propose in Section~\ref{sec:aggregation} a new aggregate
  operator on factorisations and integrate it into evaluation and
  factorisation plans. For a given aggregation function, this operator
  can reduce entire factorised relations to aggregate values and
  follows simple compositional rules.

\item We characterise in Section~\ref{sec:gby-oby} those factorisation
  trees that support efficient enumeration of result tuples for
  queries with group-by and order-by clauses. For all other
  factorisation trees, we show how to partially restructure them so as
  to enable efficient enumeration.

\item We define in Section~\ref{sec:optimisation} the search space for
  optimal evaluation plans on factorisations and introduce an
  optimisation strategy that subsumes existing techniques for eager
  and lazy aggregation~\cite{YanL:95}.

\item We extend the main-memory FDB query engine with support for
  queries with aggregates and group-by and order-by clauses.

\item We report in Section~\ref{sec:experiments} on experiments with
  FDB and the open-source relational engines SQLite and PostgreSQL.
  Our experiments confirm that the performance of these engines follow
  the succinctness gap for input data representations. Since these
  relational engines do not consider optimisations involving
  aggregates, we also report on their performance with manually
  crafted optimised query plans.

\end{compactitem}

%% file: prelims.tex
% !TEX root = main.tex

\section{Preliminaries}
\label{sec:prelims}

We assume familiarity with the vocabulary for relational
databases~\cite{AHV95}. We consider queries expressed in relational
algebra using standard operators selection, projection, join, and
additional operators for aggregation, ordering, and limit:
\begin{compactitem}
\item $\varpi_{G; \alpha_1 \leftarrow F_1, \dots, \alpha_k \leftarrow
    F_k}$ groups the input tuples by the attributes in the set $G$ and
  then applies the aggregation functions $F_1$ to $F_k$ on the tuples
  within each group; the aggregation results are labelled $\alpha_1$ to
  $\alpha_k$, respectively.

\item $o_{G}$ orders lexicographically the input relation by the list
  $G$ of attributes, where each attribute is followed by $\uparrow$ or
  $\downarrow$ for ascending or descending order, respectively; by
  default, the order is ascending and omitted.
  
\item $\lambda_k$ outputs the first $k$ input tuples in the input
  order.
\end{compactitem}

As selection conditions, we allow conjunctions of equalities $A_i =
A_j$ and $A_i \theta c$, where $A_i$ and $A_j$ are attributes, $c$ is
a constant, and $\theta$ is a binary operator. We consider standard
aggregation functions: sum, count, min, and max; avg can be seen as a
pair of (sum, count) aggregation functions.

SQL \emph{having} clauses, which are conjunctions of conditions
involving aggregate functions, attributes in the group-by list and
constants, are readily supported. Such a clause can be implemented by
adding its aggregate functions to the aggregate operator and by adding
on top of the query a selection operator whose condition is that of
the clause.

%%%%%%%%%%%%%%%%%%%%%%%%%%%%%%%%%%%%%%%%
%%%%%%%%%%%%%%%%%%%%%%%%%%%%%%%%%%%%%%%%
\subsection{Factorised Databases}

We overview necessary vocabulary on factorised databases \cite{OZ12}
and on the state of the art on the evaluation and optimisation of
select-project-join queries on such databases~\cite{BOZ12}.

{\noindent\bf Factorisations.} The key idea is to represent relations
by relational algebra expressions consisting of unions, products, and
singleton relations. A singleton relation $\data{A}{a}$ is a relation
over a schema with one attribute $A$ containing one tuple with value
$a$.

\begin{definition}
  A \emph{factorised representation (or factorisation)} over
  relational schema $\mathcal{S}$ is an expression of the form
\begin{compactitem}
\item $(E_1 \cup \dots \cup E_n)$, where each $E_i$ is a factorised
  representation over $\mathcal{S}$,
\item $(E_1 \times \dots \times E_n)$, where each $E_i$ is a
  factorised representation over $\mathcal{S}_i$ and $\mathcal{S}$ is
  the disjoint union of $\mathcal{S}_i$,
\item $\data{A}{a}$, with attribute $A$, value $a$, and $\mathcal{S} = \{A\}$,
\item $\langle\rangle$, representing the nullary tuple over
  $\mathcal{S} = \{\}$,
\item $\emptyset$, representing the empty relation over any
  $\mathcal{S}$.
\end{compactitem}
\end{definition}

Any factorisation $E$ over $\mathcal{S}$ represents a relation
$\dbracket{E}$ over $\mathcal{S}$, which can be obtained by
equivalence-preserving rewritings in relational algebra that un-nest
all unions within products. Any relation $R$ can be trivially written
as a union of products of singletons, in which each such product
corresponds to a tuple of $R$. More succinct factorisations of $R$ are
possible by the distributivity of product over union.

\begin{example}
The relation
\[R = \{(\diamondsuit,1), (\diamondsuit,2), (\diamondsuit,3), (\clubsuit,1), (\clubsuit,2), (\clubsuit,3)\}\]
over schema $\{A,B\}$ can be equivalently written as
\begin{align*}
E_1 =& (\data{A}{\diamondsuit}\!\!\times\!\!\data{B}{1}) \cup (\data{A}{\diamondsuit}\!\!\times\!\!\data{B}{2}) \cup (\data{A}{\diamondsuit}\!\!\times\!\!\data{B}{3}) \cup \\
&(\data{A}{\clubsuit}\!\!\times\!\!\data{B}{1}) \cup (\data{A}{\clubsuit}\!\!\times\!\!\data{B}{2}) \cup (\data{A}{\clubsuit}\!\!\times\!\!\data{B}{3})
\end{align*}
and can be factorised as
\[E_2 = (\data{A}{\diamondsuit} \cup \data{A}{\clubsuit}) \times (\data{B}{1} \cup \data{B}{2} \cup \data{B}{3}).\]
\par
\vspace{-1.6em}
\punto
\end{example}

Example~\ref{ex:running-example} also gives several factorisations; for
simplicity, the attributes are not shown in the singletons of these
factorisations but can be inferred from the schema of the represented
relation. The factorisations are more succinct than the represented
relations. For instance, in the factorisation in
Figure~\ref{fig:ex-db}, since the information that Lucia and
Pietro ordered Hawaii pizza on Friday is independent of the toppings
of this pizza, it is factored out and only represented once.

The tuples of the relation $\dbracket{E}$ of a factorisation $E$ can
be enumerated from $E$ with delay between successive tuples constant
in data size and linear in the schema size, which is the same as
enumerating them from $\dbracket{E}$ if it is materialised.

{\noindent\bf Factorisation trees.}  Factorisation trees act as both
sche\-mas and nesting structures of factorisations.  Several f-trees
are given in Figure~\ref{fig:running-example} and discussed in
Example~\ref{ex:running-example}.

\begin{definition}
  A \emph{factorisation tree} (f-tree for short) over a schema
  $\mathcal{S}$ is a rooted forest whose nodes are labelled by
  non-empty sets of attributes that form a partition of $\mathcal{S}$.
\end{definition}

Given a select-project-join query $Q =\pi_{\cal
  P}\sigma_\phi(R_1\times\cdots\times R_n)$, we can characterise the
f-trees that define factorisations of the query result $Q(\D)$ for any
input database $\D$. Such f-trees have nodes labelled by equivalence
classes of attributes in ${\cal P}$; the equivalence class of an
attribute $A$ consists of $A$ and of all attributes transitively equal
to $A$ in $\phi$.

\begin{proposition}[\cite{OZ12}]\label{prop:path-constraint}
  For any input database $\D$, the query result $Q(\D)$ has a
  factorised representation over an f-tree $\T$ derived from $Q$ if
  and only if $\T$ satisfies the path constraint.
\end{proposition}

The {\em path constraint} states that dependent attributes can only
label nodes along a same root-to-leaf path in $\T$.  The attributes of
a relation are dependent, since in general we cannot make any
independence assumption about the structure of a relation.  Attributes
from different relations can also be dependent. If we join two
relations, then their non-join attributes are independent conditioned
on the join attributes. If these join attributes are not in the
projection list ${\cal P}$, then the non-join attributes of these
relations become dependent. If a relation $R$ satisfies a join
dependency $\Join(X_1,X_2)$, i.e., $R =
\pi_{X_1}(R)\Join\pi_{X_2}(R)$, then the attributes in $X_1\setminus
X_2$ are independent of the attributes in $X_2\setminus X_1$ given the
join attributes $X_1\cap X_2$.

We can compute tight bounds on the size of factorisations over
f-trees~\cite{OZ12} using the notion of fractional edge cover number
of the query hypergraph~\cite{GM06}.  These bounds can be effectively
used as a cost metric for f-trees and thus for choosing a good f-tree
representing the structure of the factorised query
result~\cite{BOZ12}.  Our query optimisation approach in
Section~\ref{sec:optimisation} makes use of this cost metric.

%%%%%%%%%%%%%%%%%%%%%%%%%%%%%%
{\noindent\bf Query evaluation using f-plans.} FDB can compile any
select-project-join query into a sequence of low-level operators,
called {\em factorisation plan} or f-plan for short~\cite{BOZ12}.
These operators are mappings between factorisations. At the level of
f-trees, they can restructure by swapping parent-child nodes, merging
sibling nodes, absorbing one node into its ancestor, adding new
f-trees, and removing leaves. A product is implemented by simply
creating a forest with the two input f-trees. A selection $A_1 = A_2$
is implemented by a merge operator if the attributes $A_1$ and $A_2$
lie in sibling nodes, or by an absorb operator if $A_1$'s node is a
descendant of $A_2$'s node; otherwise, FDB swaps nodes until one of
merge or absorb operators can be applied. A projection with attributes
${\cal P}$ is implemented by removing all attributes that are not in
${\cal P}$; if this yields nodes without attributes, then
restructuring is needed so that these nodes first become leaves and
then are removed from the f-tree. A renaming operator can be used to
change the name of an attribute. In FDB, renaming needs constant time,
since the attribute names are kept in the f-tree and not with each
singleton.

\nop{We introduce one new aggregation operator in
Section~\ref{sec:aggregation}. Group-by and order-by clauses are
implemented by restructuring using the swap operator, as shown in
Section~\ref{sec:gby-oby}.}

The execution cost of f-plans is dictated by the sizes of its
intermediate and final results, which depend on the succinctness of
their factorisations. This adds a new dimension to query optimisation,
since, in addition to finding a good order of the operators, we also
need to explore the space of possible factorisations for these
results.

%% file: aggregation.tex
% !TEX root = main.tex

\section{The Aggregation Operator}
\label{sec:aggregation}

In this section we propose a new aggregation operator on factorised
data. To evaluate queries with aggregates, the FDB query engine uses
factorisation plans (sequences of operators) in which the query
aggregate is implemented by one or more aggregation operators. We next
give its semantics and linear-time algorithms that implement it.

The syntax of this operator is $\agg{F(\U)}$, where $F$ is the
aggregation function, which in our case can be any of sum, count, max,
or min (avg is recovered as a pair of sum and count), and $\U$ is a
subtree in the f-tree $\T$ of the input factorisation. In case of an
aggregation function $\mbox{sum}_A$, $\mbox{min}_A$ or $\mbox{max}_A$,
the subtree $\U$ must contain the attribute $A$.

Given a factorisation over $\T$, the operator evaluates the
aggregation function $F$ over all attributes in $\U$ and stores the
result in a new attribute $F(\U)$. Expressed as a transformation of
the relation $\dbracket{E}$ represented by the factorisation $E$, this
operator maps $\dbracket{E}$ to a relation $R =
\varpi_{\mathcal{T}\setminus\mathcal{U}; F(\U) \leftarrow F}
\dbracket{E}$ over the schema\footnote{To avoid clutter, we slightly
  abuse notation and use $\T$ to also denote the set of attributes in
  the f-tree $\T$.} $(\T\setminus\U) \cup \{F(\U)\}$.

In the resulting f-tree $\T'$, the subtree $\U$ is replaced by a new node $F(\U)$. The resulting factorisation is uniquely characterised by its underlying relation $R$ and f-tree $\T'$. Section~\ref{sub:algo} gives algorithms for our aggregation operator that computes such factorisations.

\begin{example}
\label{ex:aggregate-op}
Figure~\ref{fig:running-example} shows f-trees before and after the
execution of aggregate operators. For $F=\mbox{sum}_{\mbox{price}}$
and the subtree $\U$ rooted at node item in $\T_1$, the resulting
f-tree after the execution of the operator $\agg{F(\U)}$ is
$\T_2$. For $F = \mbox{count}_{\mbox{date}}$ and the input f-tree $\T_3$,
the resulting f-tree after the execution of the operator
$\agg{F(\mbox{date})}$ is $\T_4$.
\punto
\end{example}

Similarly to the case of select-project queries, we can characterise
precisely all f-trees $\T'$ that define the nesting structures of
factorisations for possible results of the aggregation operator
$\agg{F(\U)}$. The characterisation via the path constraint in
Proposition~\ref{prop:path-constraint} also holds for aggregation,
with the addition that the aggregation operator introduces new
dependencies among the attributes in the f-tree $\T'$.  By projecting
away the attributes in $\U$, all attributes dependent on attributes in
$\U$ now become dependent on each other (as for the projection
operator). In addition, the new attribute $F(\U)$ depends on each of
these attributes. The path constraint then stipulates that any two
dependent attributes must lie along a same root-to-leaf path in the
new f-tree $\T'$. Thus, the f-tree $\T'$ resulting from the
aggregation operator $\gamma_{F(\U)}$ satisfies the path constraint
and hence the resulting factorisation exists and is uniquely defined.

\begin{example}
Consider the aggregation operators described in Example~\ref{ex:aggregate-op}.
In $\T_2$, the only new dependency introduced by the aggregate
operator is between the new attribute $\mbox{sum}_{\mbox{price}}$ and
the attribute pizza, since we projected away the attributes item and
price that depended on the attribute pizza.

In $\T_4$, the new attribute $\mbox{count}_{\mbox{date}}$ depends on
all attributes that the attribute data depended on, namely pizza and
customer.
\punto
\end{example}

%%%%%%%%%%%%%%%%%%%%%%%%%%%%%%%%%%%%%%%%
\subsection{Composing Aggregation Operators}
\label{sec:composition}

As exemplified in the introduction, for reasons of efficiency we would
often like to execute a query by implementing aggregates via several
aggregation operators and by possibly interleaving them with
restructuring operators. This requires an approach that can compose
aggregation operators so as to implement a larger aggregate. We next
describe such an approach.

We give special status to the attributes that hold results of previous
aggregate operators, and interpret them as pre-computed values of an
aggregate instead of arbitrary data values. We refer to such
attributes as \emph{aggregate attributes}; all other attributes are
\emph{atomic}. An aggregate attribute $G(\mathcal{X})$ carries along
the aggregation function $G$ and the original attributes ${\cal X}$ to
which $G$ was applied. The aggregation operator then interprets
factorisations $\data{G(\mathcal{X})}{v}$ over the f-tree consisting
of the node $G({\cal X})$ as a relation over schema $\mathcal{X}$ and
where the aggregate value for $G$ is $v$. This special interpretation
of aggregate attributes helps us distribute the evaluation of a query
aggregate $\varpi_{G; \alpha \leftarrow F}$ over several aggregation
operators. After the last operator is executed, we execute a renaming
operator that changes the name of the last aggregation function
application to the attribute $\alpha$, as specified by the query
aggregate.

\begin{example}
  After applying the operator $\agg{count(item)}$ to the relation
  Pizzas in Figure~\ref{fig:ex-db}, we get the
  factorisation\vspace*{-.5em}

\begin{scriptsize}
\begin{align*}
&\data{\mbox{pizza}}{\mbox{Margherita}} \times \data{\mbox{count}(\mbox{item})}{1} \:\cup \\
&\data{\mbox{pizza}}{\mbox{Capricciosa}} \times \data{\mbox{count}(\mbox{item})}{3} \:\cup \\
&\data{\mbox{pizza}}{\mbox{Hawaii}} \times \data{\mbox{count}(\mbox{item})}{3}.
\end{align*}
\end{scriptsize}\vspace*{-.5em}

A subsequent $count(pizza, item)$ aggregation must interpret the
singleton $\data{\mbox{count}(\mbox{item})}{3}$ as a relation with
three items to obtain the correct result
$\data{\mbox{count}(\mbox{pizza},\mbox{item})}{7}$ and not
$\data{\mbox{count}(\mbox{pizza},\mbox{item})}{3}$. \punto
\end{example}

The composition rules for these operators are specified next using a
binary operator $\circ$: $B\circ A$ means that we first evaluate $A$
and then $B$.

\begin{proposition}\label{prop:rules}
For any (sum, count, min, max) aggregation functions $F$ and $G$ and
f-trees $\U$ and $\V$, it holds that:
\begin{itemize}
\item If $\U\supseteq\V$, then $\agg{F(\U)} \circ \agg{F(\V)} = \agg{F(\U)}$.

\item If $\U\supseteq\V$ and $A\not\in\V$, then \[\agg{\mbox{sum}_A(\U)} \circ \agg{\mbox{count}(\V)} = \agg{\mbox{sum}_A(\U)}.\]

\item If $\U\cap\V=\emptyset$, then $\agg{F(\U)} \circ \agg{G(\V)} = \agg{G(\V)} \circ \agg{F(\U)}$.
\end{itemize}
\end{proposition}

Using Proposition~\ref{prop:rules}, we can deduce that
\begin{align*}
\agg{F(\U_n)} \circ \dots \circ \agg{F(\U_1)} &= \agg{F(\U_n)}\\
\agg{F_n(\U_n)} \circ \dots \circ \agg{F_1(U_1)} &= \agg{F_n(\U_n)}
\end{align*}
for any sequence of composable aggregation operators such that
$\forall 1\leq i \leq n: \U_i\subseteq\U_n$, $F$ can be any (sum,
count, min, max) aggregation function, and $F_i$ is $sum_A$ whenever
$A\in \U_i$, and the count aggregation function otherwise. In other
words, as long as the last operator aggregates over an attribute set
$\U$, we can do pre-aggregations on subsets of $\U$.

The query aggregates can then decompose as follows: $count$ aggregates
can decompose into several $count$ operators; $sum_A$ aggregates can
decompose into a mix of $sum_A$ and $count$ operators, $min$
aggregates can decompose into several $min$ operators, and $max$
aggregates into $max$ operators.

\begin{example}
  Consider the f-tree $\T_4$ in Figure~\ref{fig:running-example} and
  the factorisation in Example~\ref{ex:running-example} that was
  obtained by executing the operators $\agg{sum_{price}(item,
    price)}$, followed by restructuring and then $\agg{count(date)}$
  on relation $R$. A subsequent operator $\agg{sum_{price}(\U)}$, with
  $\U$ the subtree rooted at node pizza, uses the results of the first
  two operators to compute the result of the query aggregate
  $\varpi_{customer; sum(price)} (R)$, as detailed in
  Example~\ref{ex:running-example}.  An alternative evaluation would
  only execute $\agg{sum_{price}(\U)}$ without executing the previous
  two aggregation operators.  We can capture this equivalence as
\begin{align*}
  & \agg{sum_{price}(\U)} \circ \agg{count(date)} \circ \agg{sum_{price}(item, price)} 
  = \agg{sum_{price}(\U)}.
\end{align*}
\end{example}

\nop{
\begin{example}
Suppose we want to evaluate the aggregate $\varpi_{\mbox{customer;
sum(price)}}$ on factorisations over the f-tree $\T_4$ from the
introduction, where we already evaluated a count aggregate on date and
a sum aggregate over price on the f-tree with item and price. What
remains to be done is to aggregate the subtree rooted at node pizza,
i.e., everything under node customer. Call this subtree $\U$ and its
two children $A$ and $B$.\nop{ (A = count_date(date) and B =
sum_price(item,price))}

For relational evaluation, we would write this aggregate as
$\varpi_{\mbox{customer; sum(A*B)}}$. The aggregation operator
$\agg{sum_{price}(\U)}$ evaluates this query by calculating $sum(A *
B)$ for each customer:

Any f-representation over $\T_4$ is of the form $\bigcup
\data{customer}{c} \times E_c$ where each E_c is an f-representation
over $U$. By definition, $\varpi_{sum_price(U)}$ replaces each E_c by
sum_price(E_c). Any f-representation over U is of the form $E_c =
\bigcap_i data{count(date)}{v_i} \times
\data{sum_price(item,price)}{u_i}$. Then by the recursive definition
of sum_price, $sum_price(E_c) = sum_price(U) = \sum_i
count(data{count(date)}{v_i}) *
sum_price(\data{sum_price(item,price)}{u_i}) = \sum_i v_i*u_i.$
\end{example}
}

%%%%%%%%%%%%%%%%%%%%%%%%%%%%%%%%%%%%%%%%
\subsection{Algorithms for the Aggregation Operator}
\label{sub:algo}

In a factorisation over an f-tree $\T$, the expressions over a subtree
$\U$ of $\T$ represent the values of the attributes of $\U$ grouped by
the remaining attributes $\T\setminus \U$. The aggregation operator
$\agg{F(\U)}$ then only replaces each such expression over $\U$ by a
singleton $\data{F(\U)}{v}$ where $v$ is the value of the aggregation
function $F$ on the relation represented by that expression.

\nop{
Let $E$ be an f-representation of a relation $R$ over an f-tree
$\T$. If $B_1$ is the root of $\T$ then $E = \bigcup_b \data{B_1}{b}
\times E_{b}$, where each $E_b$ represents
$\pi_{\T\setminus\{B_1\}}\sigma_{B_1 = b}R$. If $B_2$ is a child of
$B_1$, then each $E_b$ has a factor of the form $\bigcup_{b'}
\data{B_2}{b'} \times E_{b,b'}$, where $E_{b,b'}$ represents
$\pi_{desc(B_2)}\sigma_{B_1 = b \wedge B_2 = b'}R$ where $desc(B_2)$
denotes the descendants of $B_2$. Continuing in this manner, if $B_1,
\dots, B_k$ are the ancestors of the subtree $\U$, ordered from the
root $B_1$ of $\T$ to the direct parent $B_k$ of the root of $\U$,
then for any tuple $t$ over $\mathcal{B} = \{B_1, \dots, B_k\}$ we can
easily find the subexpression $E_t$ of $E$ which represents $\pi_\U
\sigma_{\mathcal{B} = t} R$. However, since $\U$ only depends on
$\mathcal{B}$, we have $\pi_\U \sigma_{\mathcal{B} = t} R = \pi_\U
\sigma_{\T\setminus\U = t'} R$ for any tuple $t'$ extending $t$ to all
of $\T\setminus\U$. Therefore,
\[R = \{\tuple{t, \pi_\U \sigma_{\T\setminus\U = t'} R} : t \in \pi_{\T\setminus\U}\} = \{\tuple{t, \dbracket{E_t}} : t \in \pi_{\T\setminus\U}\}.\]
The result of the operator $\varpi_{F,\U}$ is
\[R = \{\tuple{t, F(\pi_\U \sigma_{\T\setminus\U = t'} R}) : t \in \pi_{\T\setminus\U}\} = \{\tuple{t, F(\dbracket{E_t})} : t \in \pi_{\T\setminus\U}\}.\]
Therefore, it suffices to replace each $E_t$ by $F(\dbracket{E_t})$.
}

The value of $F(\dbracket{E})$ for a given factorisation $E$ over an
f-tree $\U$ can be computed recursively on the structure of $E$ in
time linear in the size of $E$, even though $E$ can be much smaller
than the relation $\dbracket{E}$ it represents. We reported in earlier
work a precise characterisation of the succinctness gap between
results to select-project-join queries and their factorisations over
f-trees~\cite{OZ12}. This gap can be exponential for a large class of
queries.

We next give algorithms for each aggregation function.

\nop{
1) how do we find/identify the expressions which correspond to the
attributes of U? (when doing operator agg{F}{U})

2) justify that to perform the operator it is enough to "cut" each
piece over U and "paste" in its place the singleton <F(U) : v> }

%%%%%%%%%%%%%%%%%%%%%%%%%%%%%%%%%%
\subsubsection{The aggregation function $count$}

We first give a recursive counting algorithm. The input is a
factorisation $E$ over an f-tree and the output is the cardinality of
the relation $\dbracket{E}$ represented by $E$.

$count(E)$:
\begin{itemize}
\item {\bf If} $E = \data{A}{a}$ for atomic attribute $A$ and value
  $a$, {\bf then}
\mbox{\bf return } $1$.

\item {\bf If} $E = \data{count(\mathcal{X})}{c}$ for any set of
  atomic attributes $\mathcal{X}$ and number $c$, {\bf then}
\mbox{\bf return } $c$.

\item {\bf If} $E = \bigcup_i E_i$, {\bf then}
\mbox{\bf return } $\textstyle\sum_i count(E_i)$,

since the relations $\dbracket{E_i}$ represented by the subexpressions
$E_i$ are disjoint.

\item {\bf If} $E = \times_i E_i$, {\bf then}
\mbox{\bf return } $\textstyle\prod_i count(E_i)$.
\end{itemize}

%%%%%%%%%%%%%%%%%%%%%%%%%%%%%%%%
\subsubsection{The aggregation function $sum_A$}

The case of a sum aggregate is similar to count. The following
algorithm $sum_A$ takes as input a factorisation $E$ over an f-tree
that contains the attribute $A$ and outputs the sum of all $A$-values
in the relation $\dbracket{E}$ represented by $E$.

$sum_A(E)$:
\begin{itemize}
\item {\bf If} $E = \data{A}{a}$, {\bf then}
  \mbox{\bf return } $a$.

\item {\bf If} $E = \data{sum_A(\mathcal{X})}{s}$ for any set of
  attributes $\mathcal{X}$ and number $s$, {\bf then}
  \mbox{\bf return } $s$.

\item {\bf If} $E = \bigcup_i E_i$, {\bf then}
\mbox{\bf return }  $\textstyle\sum_i sum_A(E_i)$,

since the relations $\dbracket{E_i}$ represented by the subexpressions
$E_i$ are disjoint.

\item {\bf If} $E = \times_i E_i$, then exactly one of the expressions 
$E_i$ has the attribute $A$ in its schema; let it be $E_j$. 

{\bf Then} \mbox{\bf return } $sum_A(E_j) * \textstyle\prod_{i\neq j} count(E_i)$.
\end{itemize}

\begin{example}
\label{ex:algo}
Consider the following factorisation\vspace*{-.5em}

\begin{scriptsize}
\begin{align*}
 \data{\mbox{customer}}{\mbox{Lucia}} &\times \data{\mbox{pizza}}{\mbox{Hawaii}}
 \times \data{\mbox{count}_{date}(\mbox{date)}}{1}  \\
 &\times \data{\mbox{sum}_{price}(\mbox{item},\mbox{price})}{9} \cup\\
 \data{\mbox{customer}}{\mbox{Mario}} &\times
 (\data{\mbox{pizza}}{\mbox{Capricciosa}} \times
 \data{\mbox{count}_{date}(\mbox{date)}}{2} \\
 &\times \data{\mbox{sum}_{price}(\mbox{item},\mbox{price})}{8} \cup \\
 \phantom{\data{\mbox{customer}}{\mbox{Mario}} \times(}&
 \hspace*{1.5em}\data{\mbox{pizza}}{\mbox{Margherita}} 
 \times \data{\mbox{count}_{date}(\mbox{date)}}{1} \\
 &\times \data{\mbox{sum}_{price}(\mbox{item},\mbox{price})}{6}) \cup\\
 \data{\mbox{customer}}{\mbox{Pietro}} &\times \data{\mbox{pizza}}{\mbox{Hawaii}}
 \times \data{\mbox{count}_{date}(\mbox{date)}}{1}  \\
 &\times \data{\mbox{sum}_{price}(\mbox{item},\mbox{price})}{9}
\end{align*}
\end{scriptsize}\vspace*{-.5em}

over the f-tree $\T_4$ from Example~\ref{ex:running-example}.  The
operator $\agg{\mbox{sum}_{price}(\U)}$, where $\U$ is the subtree of
$\T_4$ rooted at node pizza, replaces each expression over $\U$ with
the aggregate value $\mbox{sum}_{price}$ of its represented relation.
That is, we must calculate $v = \mbox{sum}_{price}
\dbracket{E}$, where\vspace*{-.5em}

\begin{scriptsize}
\begin{align*}
E = &\data{\mbox{pizza}}{\mbox{Hawaii}} \times \data{\mbox{count}_{date}(\mbox{date)}}{1} \times \\
&\data{\mbox{sum}_{price}(\mbox{item},\mbox{price})}{9}),
\end{align*}
\end{scriptsize}\vspace*{-.5em}

and replace $E$ in the factorisation by $v$:\vspace*{-.5em}

\begin{scriptsize}
\begin{align*}
\data{\mbox{customer}}{\mbox{Lucia}} \times
\data{\mbox{sum}_{price}(\mbox{item},\mbox{price}, \mbox{pizza}, \mbox{date})}{9}).
\end{align*}
\end{scriptsize}\vspace*{-.5em}

Similarly, we obtain the following for Mario and Pietro:\vspace*{-.5em}

\begin{scriptsize}
\begin{align*}
&\data{\mbox{customer}}{\mbox{Mario}} \times
\data{\mbox{sum}_{price}(\mbox{item},\mbox{price}, \mbox{pizza}, \mbox{date})}{22})\\
&\data{\mbox{customer}}{\mbox{Pietro}} \times
\data{\mbox{sum}_{price}(\mbox{item},\mbox{price}, \mbox{pizza}, \mbox{date})}{9}).
\end{align*}
\end{scriptsize}\vspace*{-.5em}

Using the algorithms, the value $v = \mbox{sum}_{price}
\dbracket{E}$ can be computed as\vspace*{-.5em}

\begin{scriptsize}
\begin{align*}
&\llbracket \data{\mbox{pizza}}{\mbox{Hawaii}} \times \data{\mbox{count}_{date}(\mbox{date)}}{1} \times
\data{\mbox{sum}_{price}(\mbox{item},\mbox{price})}{9}) \rrbracket \\
=\:& 1 \cdot \dbracket{\data{\mbox{count}_{date}(\mbox{date)}}{1}} \cdot
 \dbracket{\data{\mbox{sum}_{price}(\mbox{item},\mbox{price})}{9}} 
= 1 \cdot 1 \cdot 9 = 9.
\end{align*}
\end{scriptsize}\vspace*{-1em}

Similarly, $v = 1 \cdot ( 1 \cdot 2 \cdot 8 + 1 \cdot 1 \cdot 6) = 16
+ 6 = 22$ for Mario and $v = 1 \cdot 1 \cdot 9 = 9$ for Pietro. \punto
\end{example}

%%%%%%%%%%%%%%%%%%%%%%%%%%%%%%%%
\subsubsection{The aggregation functions $min_A$ and $max_A$}

We next give an algorithm for the aggregation function $min_A$; the
case for $max_A$ is analogous.

$min_A(E)$:
\begin{itemize}
\item {\bf If} $E = \data{A}{a}$, {\bf then} \mbox{\bf return } $a$.

\item {\bf If} $E = \data{min_A(\mathcal{X})}{c}$ for any set of attributes
  $\mathcal{X}$ and value $c$, {\bf then} \mbox{\bf return } $c$.

\item {\bf If} $E = \bigcup_i E_i$, {\bf then} \mbox{\bf return } $\textstyle\min_i min_A(E_i)$.

\item {\bf If} $E = \times_i E_i$, where $E_j$ is the expression that has
  the attribute $A$ in its schema, {\bf then} \mbox{\bf return } $min_A(E_j)$.
\end{itemize}

%%%%%%%%%%%%%%%%%%%%%%%
\subsubsection{Composite aggregation functions}
\label{sec:composite-aggregation}

For a composite aggregation function $(F,G)$, such as $avg_A $ $=
(sum_A,count)$, we apply the algorithms for the constituent
aggregation functions $F$ and $G$ separately. For an input
factorisation $E$, we then obtain $(F(E),G(E))$, e.g., $(sum_A(E),$
$count(E))$ in case of $avg_A$. 

Query aggregates with more than one aggregation function also call for
composite aggregate functions. For instance, the query aggregate
$\varpi_{G; \alpha_1 \leftarrow F_1, \dots, \alpha_k \leftarrow F_k}$
require the evaluation of a $k$-ary aggregation function
$F=(F_1,\ldots,F_k)$. As for unary aggregates, the evaluation of
composite aggregates can be distributed over several aggregation
operators. Since the grouping attributes $G$ are the same for all
aggregation functions in $F$, each of these operators aggregates over
the same f-tree for all aggregation functions in $F$. The resulting
singletons in the factorisation would have the form
$\data{(F_1,\ldots,F_k)}{(v_1,\ldots,v_k)}$, where $v_i$ would be the
result of applying the aggregation $F_i$ on the input.

If the same aggregation function has to be applied several times, we
calculate its result value only once. This situation arises e.g.\ in
case of the $avg_A$ aggregate or more generally for query aggregates
with $count$ and $sum_A$ functions, since $sum_A$ is decomposed into
$sum_A$ and $count$, and the two $count$ computations can be shared.

%% file: gby-oby.tex
% !TEX root = main.tex

\section{Group-by and Order-by Clauses}
\label{sec:gby-oby}

We next address the problem of evaluating group-by and order-by
clauses on factorised data.  While these query constructs do not
change the data, they may restructure it.

On relational data, grouping by a set $G$ of attributes partitions the
input tuples into groups that agree on the value for $G$. Grouping is
solely used in connection with aggregates, where a set of aggregates
are applied on the tuples within each group.  One approach to
implementing grouping is to sort the input relation on the attributes
of $G$ using some order of these attributes; this is similar in spirit
to the approach taken by the FDB.

Ordering an input relation by a list $O$ of attributes sorts the input
relation lexicographically on the attributes in the order given by
$O$; for each attribute in $O$, we can specify whether the sorting is
in ascending or descending order.

The tuples in the sorted relation, or within a group in case of
grouping, can then be enumerated in the desired order with {\em
constant delay}, i.e., the time between listing a tuple and listing
its next tuple in the desired order is constant and thus independent
of the number of tuples. The limit query operator $\lambda_k$ then
only returns the first $k$ tuples from the sorted relation.

In case of factorisations, tuple enumeration in a given order or by
groups may require restructuring. This restructuring task can be
effected without the need to flatten the factorisations. In the
following, we first characterise those factorisations that support
constant-delay enumeration in a given order, and then explain how to
restructure all other factorisations to meet the constraint.

%%%%%%%%%%%%%%%%%%%%%%%%%%%%%%%%%%%%%%%%%%%
%%%%%%%%%%%%%%%%%%%%%%%%%%%%%%%%%%%%%%%%%%%
\subsection{F-tree Characterisation by Constant-Delay Enumeration in
  Given Orders}

For any factorisation $E$ over an f-tree $\T$, it is possible to
enumerate\footnote{The enumeration procedure uses a hierarchy of
iterators in the parse tree of the factorisation, one per node in
$\T$.}  the tuples in the represented relation $\dbracket{E}$ in {\em
no particular order} with constant delay~\cite{BOZ12}; more precisely,
the delay is linear in the size of the schema, which is fixed.

The goal of this section is to characterise those f-trees $\T$
defining factorisations for which constant-delay enumeration also
exists for some given orders.

For any attribute, its singletons within each union are kept sorted in
ascending order and all operators preserve this ordering constraint.
This holds for the FDB implementation and also for all example
factorisations in this paper. This sorting is used for efficient
implementation of equality selections as intersection of sorted lists.
It also serves well our enumeration purpose. In particular, any
factorisation already supports constant-delay enumeration in certain
orders, for example those representing prefixes of paths in the f-tree
of the factorisation.

\begin{example}
  The f-tree $\T_1$ in Figure~\ref{fig:running-example} supports
  constant-delay enumeration in any of the orders (pizza); (pizza,
  date); (pizza, date, customer); (pizza, item); or (pizza, item,
  price); (pizza, date, item); but not in the orders (pizza, customer,
  date); (customer, pizza). This can be verified on the factorisation
  over $\T_1$ given in Figure~\ref{fig:ex-db}. The order on each of
  these attributes need not be ascending. Indeed, if for instance the
  order on the pizza attribute is descending, we iterate on the sorted
  list of pizzas from the end of the list to the front.\punto
\end{example}

In contrast to ordering, grouping is less restrictive since the order
of the attributes in the group is not relevant. An f-tree then readily
supports constant-delay enumeration of tuples by groups in a larger
number of orders.

\begin{example}
  The f-tree $\T_1$ in Figure~\ref{fig:running-example} supports
  constant-delay enumeration for grouping over all orders mentioned in
  the previous example as well as all their permutations.\punto
\end{example}

We next make this intuition more precise. For the following
statements, we assume without loss of generality that no two
attributes in the attribute group $G$ or order list $O$ are within the
same equivalence class; if they are, their values are the same for
each tuple and we can ignore one of these attributes in $G$ and the
last of the two in the ordered list $O$.

\begin{theorem}\label{th:group-by}
  Given a factorisation $E$ over an f-tree $\T$ and a set $G$ of
  group-by attributes, the tuples within each group of $\dbracket{E}$
  can be enumerated with constant delay if and only if each attribute
  of $G$ is either a root in $\T$ or a child of another attribute of
  $G$.
\end{theorem}

The case of order-by clauses is more restrictive.

\begin{theorem}\label{th:order-by}
  Given a factorisation $E$ over an f-tree $\T$ and a list $O$ of
  order-by attributes, the tuples in $\dbracket{E}$ can be enumerated
  with constant delay in sorted lexicographic order by $O$ if and only
  if each attribute $X$ of $O$ is either a root in $\T$ or a child of
  an attribute appearing before $X$ in $O$.
\end{theorem}

\nop{The proofs of these theorems are available online~\cite{Tomas2012}.}

%%%%%%%%%%%%%%%%%%%%%%%%%%%%%%%%%%%%%%%%%%%
%%%%%%%%%%%%%%%%%%%%%%%%%%%%%%%%%%%%%%%%%%%
\subsection{Restructuring Factorisations for Order-by and Group-by Clauses}

Restructuring factorisations can be implemented using the swap
operator~\cite{BOZ12}. We first recall this operator and then discuss
how it can be effectively used to implement group-by and order-by
clauses.

Given an f-tree $\T$, the swap operator $\chi_{\A,\B}$ exchanges a
node $\B$ with its parent node $\A$ in $\T$ while preserving the path
constraint. We promote $\B$ to be the parent of $\A$ and move up its
children that do not depend on $\A$. The effect of the swapping
operator $\chi_{A,B}$ on the relevant fragment of $\T$ is shown below,
where $\T_\B$ and $\T_{\A\B}$ denote the collections of children of
$\B$ that do not depend, and respectively depend, on $\A$, and $\T_A$
denotes the subtree under $\A$. Separate treatment of the subtrees
$\T_\B$ and $\T_{\A\B}$ is required so as to preserve the path
constraint. The resulting f-tree has the same nodes as $\T$ and the
represented relation remains unchanged:

\begin{small}
\[
\psset{levelsep=5mm, nodesep=2pt, treesep=5mm}
\pstree{\TR{\cdots}}
{
  \pstree{\TR{\A}}
  {
    \TR{\T_\A}
    \pstree{\TR{\B}}
    {
      \TR{\T_{\A\B}}
      \TR{\T_\B}
    }
  }
}
\mapsto
\pstree{\TR{\cdots}}
{
  \pstree{\TR{\B}}
  {
    \pstree{\TR{\A}}
    {
      \TR{\T_\A}
      \TR{\T_{\A\B}}
    }
    \TR{\T_\B}
  }
}
\]
\end{small}

While the above explanation of the swap operator was given in terms of
f-tree manipulation, the operator also restructures factorisations
over this f-tree $\T$. Any factorisation over the relevant part of the
input f-tree $\T$ has the form
\[\textstyle\bigcup_{a} \left( \data{\A}{a} \times E_a \times \bigcup_{b} \left(\data{B}{b} \times F_b \times G_{ab} \right)\right),\]
while the corresponding restructured factorisation is
\[\textstyle\bigcup_{b} \left( \data{\B}{b} \times  F_{b} \times \bigcup_{a} \left(\data{\A}{a} \times E_{a} \times G_{ab} \right)\right).\]
The expressions $E_a$, $F_b$ and $G_{ab}$ denote the factorisations
over the subtrees $\T_\A$, $\T_\B$ and respectively $\T_{\A\B}$.  The
swap operator $\chi_{\A,\B}$ thus takes an f-representation where data
is grouped first by $\A$ then $\B$, and produces an f-representation
grouped by $\B$ then $\A$.

To restructure any f-tree $\T$ so that constant-delay enumeration is
enabled (i) for grouping by a set $G$ of attributes and then (ii) for
ordering by a list $O$ of attributes, we essentially follow the
characterisation of good f-trees given by Theorems~\ref{th:group-by}
and~\ref{th:order-by}. For grouping, we push all attributes in $G$
above all other attributes. For ordering, we proceed as for grouping
and in addition ensure that the attribute order in the list $O$ does
not contradict the root-to-leaf order in the f-tree. Each attribute
push can be implemented by a swap operator. The actual order of the
swap operators can influence performance and is thus subject to
optimisation which is discussed in Section~\ref{sec:optimisation} in
the greater context of optimisation of f-plans with several other
operators.

%% file: optimisation.tex
%!TEX root = main.tex

%%%%%%%%%%%%%%%%%%%%%%%%%%%%%%%%%%
\begin{figure*}[t]
\begin{small}
$$
\underbrace{
\begin{array}{l}
R_1 = \mbox{Orders}\Join\mbox{Items}\Join\mbox{Packages}\\
Q_1 =\varpi_{\mbox{package, date, customer;  sum(price)}} (R_1)\\
Q_2 =\varpi_{\mbox{customer; revenue $\leftarrow$ sum(price)}} (R_1)\\
Q_3 =\varpi_{\mbox{date, package;  sum(price)}} (R_1)\\
Q_4 =\varpi_{\mbox{package;  sum(price)}} (R_1)\\
Q_5 =\varpi_{\mbox{sum(price)}} (R_1)
\end{array}
}_{\mbox{AGG}}
\hspace*{.1em}%
\underbrace{
\begin{array}{l}
\ \\
\ \\
\ \\
Q_6 =o_{\mbox{customer}}(Q_2)\\
Q_7 =o_{\mbox{revenue}}(Q_2)\\
Q_8 =o_{\mbox{date, package}}(Q_3)\\
Q_9 =o_{\mbox{package, date}}(Q_3)
\end{array}
}_{\mbox{AGG+ORD}}
\hspace*{.1em}%
\underbrace{
\begin{array}{l}
R_2 = o_{\mbox{package, date, item}}(R_1)\\
R_3 = o_{\mbox{date, customer, package}}(\mbox{Orders})\\
Q_{10} = R_2\\
Q_{11} = o_{\mbox{package, item, date}}(R_2)\\
Q_{12} = o_{\mbox{date, package, item}}(R_2)\\
Q_{13} = o_{\mbox{customer, date, package, item}}(R_3)
\end{array}
}_{\mbox{ORD}}
$$
\end{small}\vspace*{-1em}

\caption{Three sets of queries used in the experiments.  Relations $R_1$, $R_2$,
  and $R_3$ are materialised.}
\label{fig:queries}
\vspace*{-1em}
\end{figure*}
%%%%%%%%%%%%%%%%%%%%%%%%%%%%%%%%%%

\section{Query Optimisation}
\label{sec:optimisation}

FDB compiles a query into a sequence of operators, called an f-plan.
While in Section~\ref{sec:composition} we present rules for composing
aggregation operators, in this section we define possible f-plans for
arbitrary queries with aggregates and present algorithms for finding
f-plans. There exist several f-plans for a given query that differ in
the join order, sequence of partial aggregates, and factorisation
restructuring.

%%%%%%%%%%%%%%%%%%%%%%%%%%%%%%%%%%%%
%%%%%%%%%%%%%%%%%%%%%%%%%%%%%%%%%%%%

\begin{example}
Consider the aggregate operator
\[ \varpi_{\mbox{customer; revenue $\leftarrow$ sum(price)}} \] 

whose input factorisation is over the f-tree $\T_1$.
Example~\ref{ex:running-example} describes an f-plan for this
query. It computes $\mbox{sum}_{price}(\mbox{item},$ $\mbox{price})$
to obtain a factorisation over $\T_2$, then pushes the node
$\mbox{customer}$ to the root and aggregates the remaining
attributes. Assuming that $\mbox{pizza}$ and $\mbox{customer}$ are
independent (e.g. if the relation Orders(pizza, date, customer) was
obtained as a join of the daily Menu(pizza, date) and Guests(date,
customer)), a different plan executes the same query. It would also
first compute $\mbox{sum}_{price}(\mbox{item}, \mbox{price})$ and
obtain a factorisation over $\T_2$. Then, it would push the node
$\mbox{date}$ to the root using a swap operator:
\begin{small}
\[
\psset{levelsep=5mm, nodesep=2pt, treesep=5mm}
\pstree{\TR{\mbox{date}}}
{
  \TR{\mbox{customer}}
  \TR{\mbox{pizza}}{
    \TR{\mbox{sum}_{\mbox{price}}(\mbox{item, price})}
  }
}
\]
\end{small}
The next operator would aggregate its two rightmost children to obtain a
factorisation with structure
\begin{small}
\[
\psset{levelsep=5mm, nodesep=2pt, treesep=5mm}
\pstree{\TR{\mbox{date}}}
{
  \TR{\mbox{customer}}
  \TR{\mbox{sum}_{\mbox{price}}(\mbox{item, pizza, price})}
}
\]
\end{small}
and only then swap $\mbox{customer}$ to the root to restructure the
factorisation as follows:
\begin{small}
\[
\psset{levelsep=5mm, nodesep=2pt, treesep=5mm}
\pstree{\TR{\mbox{customer}}}
{
  \pstree{\TR{\mbox{date}}}{
    \TR{\mbox{sum}_{\mbox{price}}(\mbox{item, pizza, price})}
  }
}
\]
\end{small}
We next perform the final aggregate and obtain 
\begin{small}
\[
\psset{levelsep=5mm, nodesep=2pt, treesep=5mm}
\pstree{\TR{\mbox{customer}}}
{
    \TR{\mbox{sum}_{\mbox{price}}(\mbox{item, pizza, price, date})}
}
\]
\end{small}
The last operator renames $\mbox{sum}_{\mbox{price}}(\mbox{item,
  pizza, price, date})$ to revenue.  \punto
\end{example}

The goal of query optimisation is to find an f-plan whose execution
time is low. The cost metric that we use to differentiate between
plans is based on asymptotically tight upper bounds on the sizes of
the factorisations representing intermediate and final results. Size
bounds are a good prediction for the time needed to create such
factorisations. As shown in earlier work~\cite{BOZ12}, this can be
computed by inspecting the f-tree of each of these results as well as
using the sizes of the input relations.

We next qualify which sequences of operators correctly execute the
query. Then, we describe two optimisation techniques: an exhaustive
search in the space of all possible operators that finds the cheapest
f-plan under a given metric but requires exponential time in the query
size, and a greedy heuristic whose running time is polynomial. Both
techniques subsume the respective optimisation techniques given in
earlier work for select-project-join queries~\cite{BOZ12}.

%%%%%%%%%%%%%%%%%%%%%%%%%%%%%%%%%%%%%
\subsection{Search Space of F-plans for a Given Query}

We consider a general\footnote{This discussion can be extended to
  composed aggregation functions following our remarks from
  Section~\ref{sec:composite-aggregation}.} query with ordering,
aggregates and selections of the form
\[Q = o_L \big( \varpi_{G; \alpha \leftarrow F} (\sigma_{A_1=B_1,
  \dots, A_m=B_m,\phi} (R_1\times\cdots\times R_n))\big)\]

Since product operators are the cheapest operators to execute on
factorisations, we always execute them first: a product of $n$
relations can be represented as a factorisation that is a product
relational expression whose children are the $n$ relations. Selections
with constants expressed by the condition $\phi$ can also be evaluated
in one traversal of this factorisation. The remaining query constructs
can be implemented using further f-plan operators; a list of available
f-plan operators is given in Section~\ref{sec:prelims}, with the
addition of the newly introduced aggregation operator.

Let $Q$ be a query without products or selections with constants, and
let $E$ be a factorisation over an f-tree $\T$. A sequence of
operators $S$ correctly implements the query $Q$ on $E$ if and only if
it satisfies the following three conditions:
\begin{compactitem}
\item \nop{(selection)} For each selection  condition $A_i =
  B_i$ in $Q$, $S$ contains a selection operator that merges the
  equivalence classes of $A_i$ and $B_i$ as well a their nodes in its
  input f-tree. No selection operator in $S$ merges nodes with
  attributes that are not equivalent in the selection of $Q$ or in the
  original f-tree $\T$.

\item \nop{(aggregation)} The sequence $S$ contains the
  aggregation operator $\agg{F(\U)}$ for some subtree $\U$ whose set
  of attributes is $\T \setminus G$. It may be preceded by any number
  of aggregates $\agg{F(\V)}$ with $\V \subseteq (\T \setminus G)$ to
  implement partial aggregation, as allowed by the composition rules
  of Proposition~\ref{prop:rules}. There are no other aggregation
  operators.  A renaming operator occurs after $\agg{F(\U)}$ in $S$ to
  rename $F(\U)$ to $\alpha$.

\item \nop{(order)} The output f-tree of the last operator must
  satisfy the condition of Theorem~\ref{th:order-by} for the order-by
  list $L$.
\end{compactitem}

These conditions present global requirements on an f-plan: they
characterise which sequences of operators correctly execute a given
query $Q$. Next we turn them into local requirements: at any point in
the f-plan we characterise which single operator can be evaluated next
so that we can still arrive at the result of $Q$. This is possible
since at any stage of the f-plan, the f-tree encodes information about
the previous operators in the f-plan as well as about the structure of
the factorisation. The nodes of the f-tree encode information about
the underlying relation (equalities and aggregates already performed),
and the shape of the f-tree encodes information about how the relation
is factorised.

Consider the scenario of executing a query $Q$ on an input
factorisation over the f-tree $\T$, and suppose we already executed a
sequence $S'$ of operators. Call an operator \emph{permissible} if it
is one of the following:

\begin{compactitem}

\nop{
\item Any selection operator merging nodes $\A$ and $\B$ that contain
  attributes equivalent in $Q$.(Recall that there are two kinds of
  selection operators for factorisations; $\mu_{\A,\B}$ can only merge
  siblings $\A$ and $\B$ and $\alpha_{\A,\B}$ merges $\B$ into an
  ancestor $\A$.)
}

\item Any selection operator for one of the remaining selections
  $A_i=B_i$ to be executed; we consider two selection operators, one
  operator (merge) requires the attributes $A_i$ and $B_i$ to be
  siblings in the f-tree, the other operator (absorb) requires one of
  the attributes to be a descendant of the other in the f-tree.

\item Any aggregate operator $\agg{F(U)}$ with $U \subseteq (\T
  \setminus G)$ is permissible unless $U$ contains an attribute $A_i$
  that is still to be equated with $B_i$. (Otherwise the equality $A_i
  = B_i$ could not be performed afterwards.)

\item Any restructuring (swap) operator.
\end{compactitem}

\begin{proposition}
\label{prop:local}
The sequence $S'$ followed by the operator $x$ can be extended to an
f-plan of $Q$ if and only if $x$ is permissible.
\end{proposition}

We can represent the space of all f-plans as a graph whose nodes are
f-trees and whose edges are operators between them. An f-plan then
corresponds to a path in the graph, and an f-plan for the query $Q$ is
a path to any f-tree satisfying the selection, aggregation, and order
conditions. In the presence of a cost metric for individual
operators, such as the one based on size bounds for factorisations of
operator outputs, we can utilise Dijkstra's algorithm to find the
minimum-cost f-plan executing the query
$Q$. Proposition~\ref{prop:local} characterises the outgoing edges for
each node and allows us to construct the graph incrementally as it is
explored.

%%%%%%%%%%%%%%%%%%%%%%%%%%%%%%%%%
\begin{figure}[t]
%  \hspace{-1em}
  \includegraphics{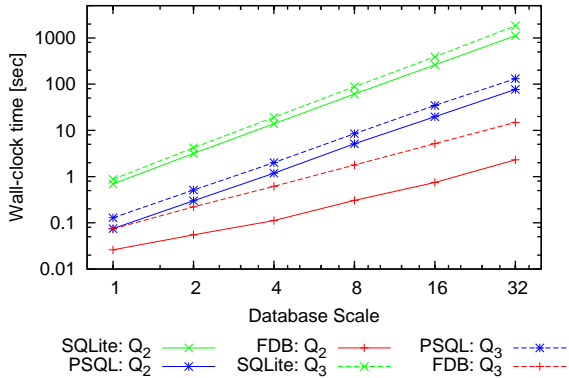}
  \vspace*{-1em}

  \caption{The effect of dataset scale on performance.}
  \label{fig:exp-scale}
  \vspace*{-1em}
\end{figure}
%%%%%%%%%%%%%%%%%%%%%%%%%%%%%%%%%

%%%%%%%%%%%%%%%%%%%%%%%%%%%%%%
\subsection{Greedy Heuristic}

The size of the space of all f-plans is exponential in the size of the
query, and searching for the optimal f-plan becomes impractical even
for simple queries. We propose a polynomial-time greedy heuristic
algorithm for finding an f-plan for a given query $Q$ and input f-tree
$\T_0$:

\medskip
Repeat
\begin{compactenum}
\item If there are any permissible selection operators, choose one
  involving a highest-placed node in the f-tree and execute it.

\item Else if there are any permissible aggregate operators
  $\agg{F(U)}$, choose one with maximal $\U$ and execute it.

\item Else if there still exists a condition $A_i=B_i$ such that $A_i$
  and $B_i$ are not in the same node, calculate the cost for
  repeatedly pushing up (a) $A_i$, or (b) $B_i$, or (c) both $A_i$ and
  $B_i$, until $A_i$ and $B_i$ are siblings or one is an ancestor of
  the other.  Pick the cheapest option and execute it.
\item Else if there is an attribute $A \in G$ with parent $B \notin
  G$, swap $A$ with $B$.
\item Else if there is an attribute $A \in L$ with parent $B$ such
  that $B$ is not before $A$ in $L$, swap $A$ with $B$,
\item Else break.

\end{compactenum}
\medskip

After this algorithm terminates, all selection conditions have been
evaluated in (1) possibly using the restructuring from (3), all
attributes not in $G$ have been aggregated in (2) possibly using the
restructuring in (4), and the order condition is met because of (5).
There may still be several aggregate attributes in the f-tree, the
value of the final aggregate is the product (or min or max, depending
on the aggregation function $F$) of these values. This can be
calculated during enumeration.

If we require the result of the aggregate in a single attribute, we
need to arrange all nodes dependent on the aggregation attributes into
a single path. This can be achieved by repeatedly swapping them up:

\medskip
\begin{compactenum}
\item[7.] Let $P$ be the least common ancestor in $\T$ of all
  attributes of $\T\setminus G$. While $P$ has a child with an atomic
  attribute $R$, swap $P$ and $R$.
\end{compactenum}

%\todo{The above says nothing about how the agg optimisations are
%  considered here.}

\nop{
A detailed evaluation of the heuristic algorithm for order-by and
group-by queries is available in a technical report~\cite{Tomas2012}.
}

%% file: experiments.tex
% !TEX root = main.tex

%%%%%%%%%%%%%%%%%%%%%%%%%%%%%%%%%%%%%%%%
\section{Experimental Evaluation}
\label{sec:experiments}

%%%%%%%%%%%%%%%%%%%%%%%%%%%%%%%%%
\begin{figure}[t]
%  \hspace{-1em}
  \includegraphics[scale=1.4]{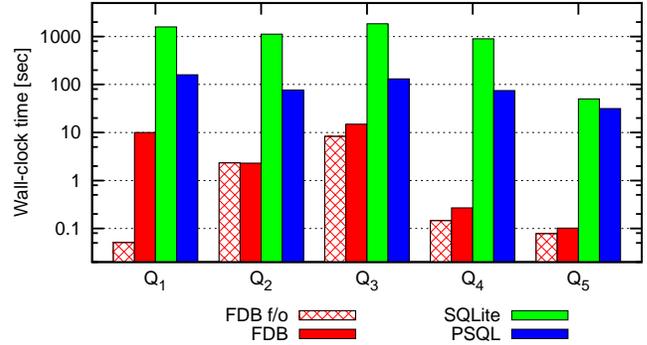}
  \caption{Performance of AGG queries on the (factorised) materialised
    view $R_1$ at scale 32.}
  \label{fig:exp-view}
  \vspace*{-1em}
\end{figure}
%%%%%%%%%%%%%%%%%%%%%%%%%%%%%%%%%

We evaluate the performance of our query engine FDB against the SQLite
and Post\-gre\-SQL open-source relational engines. Our main finding is
that FDB outperforms relational engines if its input is factorised.
In addition to the gain brought by factorisations, two optimisations
supported by FDB are particularly important: (1) Partial aggregation
that reduces the size of intermediate factorisations and (2) the reuse
of existing sorting orders as enabled by local, partial restructuring
of factorisations.  While the former optimisation is relevant to
queries with aggregates, the second is essential to queries with
order-by clauses and can make a difference even for simple queries
that just sort the input relation.  We also found that {\em limit}
clauses, which allow users to ask for the first $k$ tuples in the
result, can benefit from factorisations coupled with partial
restructuring.

{\noindent\bf Competing Engines.} FDB is implemented in C++ for
execution in main memory. We consider two flavours in the experiments:
{\bf FDB} produces flat, relational output, whereas {\bf FDB f/o}
produces factorised output. The lightweight query engine {\bf SQLite}
3.7.7 was tuned for main memory operation by turning off the journal
mode and synchronisations and by instructing it to use in-memory
temporary store. Similarly, we run PostgreSQL 9.1.8 ({\bf PSQL}) with
the following parameters: fsync, synchronous commit, full page
writes and background writer are off, shared buffers, working memory
and effective cache size increased to 12 GB. For PSQL we run each
query three times and time the last repetition, for which internal
tables are cached and queries are optimised for main memory.
For all engines we report wall-clock times to execute the query plans; these
times exclude importing the data from files on disk and writing the
result to disk. Our measurements indicate that the disk I/O for
SQLite is zero and zero or negligible for PSQL (always smaller than
when reading input and writing output to disk, which increases
execution time by at most 10\%.)

{\noindent\bf Experimental Setup.} All experiments were performed on
an Intel(R) Xeon(R) X5650 dual 2.67GHz/64bit/59GB running VMWare VM
with Linux 3.0.0/gcc4.6.1.

%%%%%%%%%%%%%%%%%%%%%%%%%%%%%%%%%
\begin{figure}[t]
%  \hspace{-1em}
  \includegraphics[scale=1.4]{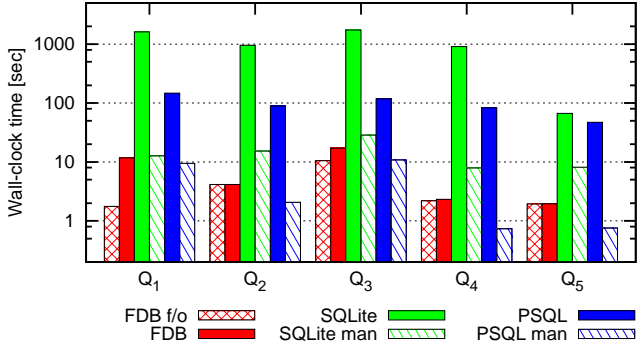}  \vspace*{-1em}

  \caption{Performance of AGG queries on flat input at scale 32.
  SQLite and PSQL using their own plans and also manually optimised
  plans (man).}  \label{fig:exp-query} \vspace*{-1em}
\end{figure}
%%%%%%%%%%%%%%%%%%%%%%%%%%%%%%%%%

%%%%%%%%%%%%%%%%%%%%%%%%%%%%%%%%%%%%%%%%
{\noindent\bf Experimental Design.}  We use a synthetic dataset that
consists of three relations: Orders, Items, Packages, generalising the
pizzeria data\-base from Example~\ref{ex:running-example}. We control
their sizes, and therefore the succinctness gap between factorised and
flat results of queries on this dataset, using a scale parameter
$s$. The number of dates on which orders are placed is $800s$, the
average number of order dates per customer is $80s$ and the average
number of orders per order date is 2, both with a binomial
distribution. There are $100\sqrt{s}$ different items and $40\sqrt{s}$
packages of $20\sqrt{s}$ items in average. Scaling generates database
instances for which the size of the natural join of all three
relations grows as $s^4$ while its factorisation over the following f-tree 
$\T$ grows as $s^3$.
\vspace*{-1em}
\[
\psset{levelsep=5mm, nodesep=2pt, treesep=5mm}
\pstree{\TR{\mbox{package}}}
{
  \pstree{\TR{\mbox{date}}}
  {
    \TR{\mbox{customer}}
  }
  \pstree{\TR{\mbox{item}}}
  {
    \TR{\mbox{price}}
  }
}
\]
For the scale factor 32, the join has 280M tuples (1.4G singletons),
while the factorisation has 4.2M singletons.

\nop{We control
its size using the scale parameter $s$: There are $100 \cdot s$ items,
$40 \cdot s$ packages and $12800\cdot s^2$ orders. \nop{We model a
  realistic scenario by scaling the data statistics as follows.} The
number of dates scales linearly with $s$, the number of customers
scales with $\sqrt{s}$. The number of dates each customer places an
order follows a binomial distribution with mean proportional to
$\sqrt{s}$. The average number of placed orders on such days is 2 and
the ordered packages are chosen uniformly at random.

The natural join $R_1$ of all relations has 280M tuples which makes 1.4G
singletons for scale factor 32. The size of the factorisation of this
join relation over the f-tree $\T = $\vspace*{-1em}
\[
\psset{levelsep=5mm, nodesep=2pt, treesep=5mm}
\pstree{\TR{\mbox{package}}}
{
  \pstree{\TR{\mbox{date}}}
  {
    \TR{\mbox{customer}}
  }
  \pstree{\TR{\mbox{item}}}
  {
    \TR{\mbox{price}}
  }
}
\]
is 4.2M singletons. By scaling the database, the factorisation size
grows as $s^3$ while the join size grows as $s^4$.
}

We use three sets of queries, cf.\@ Figure~\ref{fig:queries}.  The set
AGG consists of five queries with aggregates and group-by clauses\nop{
that are evaluated on the materialised view represented by the natural
join $R_1$, which in the case of FDB is factorised using $\T$}. The
set AGG+ORD consists of four queries with order-by clauses and
aggregates.  The set ORD consists of four order-by queries on top of
sorted relations $R_2$ and $R_3$.

We next present five experiments whose focus is on performance of
query evaluation; a comprehensive experimental evaluation of our query
optimisation techniques is available online at the FDB web page:
\begin{small}\url{http://www.cs.ox.ac.uk/projects/FDB/}\end{small}.  
For all queries used in the following experiments, the heuristic
algorithm gives optimal f-plans under the asymptotic bounds metric.

\nop{
and we only report it briefly here. We generated at random a
set of queries with $1\leq k\leq 9$ join conditions on 4 relations
with 10 attributes, and $1\leq m\leq 9$ attributes in the order-by and
group-by clauses. We found that our heuristic performs under 0.1
seconds and can be up to two orders of magnitude faster than
exhaustive search while preserving the accuracy of the latter.
}

%%%%%%%%%%%%%%%%%%%%%%%%%%%%%%%%%%%%%%%%
{\noindent\bf Experiment 1: Aggregate queries on materialised views.}
Figure~\ref{fig:exp-scale} shows that FDB clearly outperforms SQLite
and PSQL in our experiments on the factorised materialised view with
AGG queries $Q_2$ and $Q_3$. The evaluation of both queries in FDB is
done using partial aggregation and restructuring, similar to the query
$P$ in the introduction. The relational engines only perform grouping
and aggregation on the materialised view; PSQL uses hashing while
SQLite uses sorting to implement grouping. The performance gap widens
as we increase the scale factor and raises from one order of magnitude
for scale 1 to two orders of magnitude for scale 32 when compared to
PSQL; SQLite shows one additional order of magnitude gap. Notably, the
reported timing for FDB includes the enumeration of result tuples,
i.e., its output is flat as for relational engines.

Figure~\ref{fig:exp-view} looks closer at this scenario for scale 32
and AGG queries. When computing the result as factorised data, the
performance gap further widens by two orders of magnitude for $Q_1$.
This is the time needed to enumerate the tuples in the query result
and is directly impacted by the cardinality of the result. The result
of $Q_1$ is large as it consists of all joinable triples of packages,
dates, and customers. The results of the other queries have less
attributes and smaller sizes. For them, the enumeration time takes
comparable or much less time as computing the factorised result.

\nop{
In both figures, PSQL consistently outperforms SQLite with one
exception: $Q_5$ in Figure~\ref{fig:exp-view}. This query aggregates
over the whole relation and SQLite's table scanning procedure is
faster in this case.
}

%%%%%%%%%%%%%%%%%%%%%%%%%%%%%%%%%%%%%%%%
{\noindent\bf Experiment 2: Aggregate queries on relational data.}
Figure~\ref{fig:exp-query} presents FDB's performance for evaluating
aggregate queries on flat, relational data (no materialised view this
time) and producing flat output.  Surprisingly, FDB outperforms SQLite
and PSQL in their domain. A closer look revealed that both relational
engines do not use partial aggregation and hence only consider
sub-optimal query plans. With handcrafted plans that make use of
partial and eager aggregation~\cite{YanL:95}, all engines perform
similarly. If we set for factorised output, then FDB f/o outperforms
FDB in case of large factorisable results ($Q_1, Q_3$); for small
results, there is no difference to FDB as expected.

%%%%%%%%%%%%%%%%%%%%%%%%%%%%%%%%%
\begin{figure}[t]
%  \hspace{-1em}
  \includegraphics[scale=1.4]{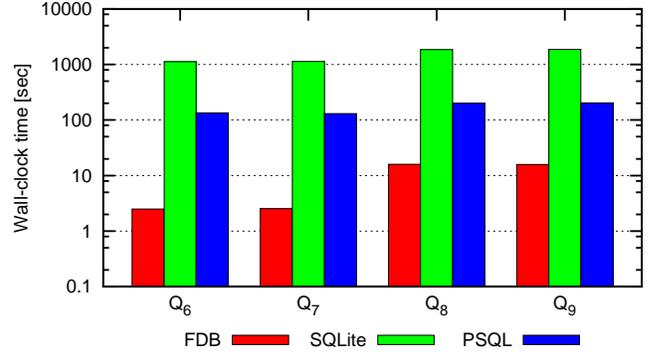}
  \caption{Performance of AGG+ORD queries on the (factorised)
    materialised view $R_1$ at scale 32. \nop{The output is flat for all
    engines.}}
  \label{fig:exp-orderby}
  \vspace*{-1em}
\end{figure}
%%%%%%%%%%%%%%%%%%%%%%%%%%%%%%%%%

%%%%%%%%%%%%%%%%%%%%%%%%%%%%%%%%%%%%%%%%
{\noindent\bf Experiment 3: Aggregate and order-by queries on
  materialised views.} Figure~\ref{fig:exp-orderby} shows that
ordering only adds a small overhead to queries with aggregates.  For
FDB, the result of $Q_2$ is already ordered by customer, and thus the
additional order-by clause in $Q_6$ is simply ignored by FDB.
Re-ordering by the result of aggregation, as done in $Q_7$, does only
add a marginal overhead, not visible in the plot due to the log scale
on the y axis. This is explained by the relatively small result of
$Q_2$. A similar situation is witnessed for the pair of queries $Q_8$
and $Q_9$ that apply different orders on the result to $Q_3$. Overall,
it takes longer since $Q_3$ has a larger result than $Q_2$ (there are
more pairs of date and package than customers). \nop{There is still
  almost no visible performance difference between $Q_8$ and $Q_9$.}
Following the pattern for queries $Q_2$ and $Q_3$ discussed in
Experiment 1 and the lack of impact of ordering in this experiment,
FDB outperforms the relational engines in this experiment, too.

\begin{figure}[t]
%  \hspace{-1em}
  \includegraphics[scale=1.4]{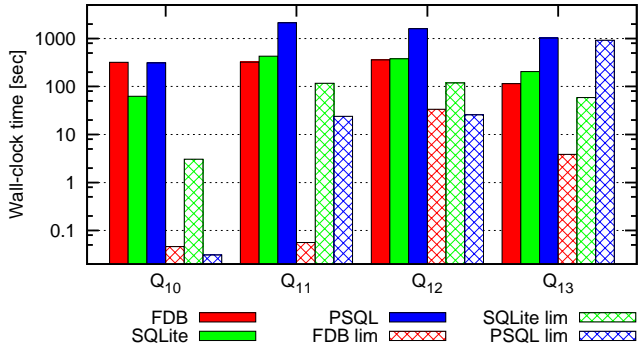}
  \caption{Performance of ORD queries with and without a LIMIT 10 statement
    (lim) on the (factorised) materialised view $R_1$ at scale 16.}
  \label{fig:exp-sort}
  \vspace*{-1em}
\end{figure}

%%%%%%%%%%%%%%%%%%%%%%%%%%%%%%%%%%%%%%%%
{\noindent\bf Experiment 4: Partial sorting via restructuring of
  factorisations.} In this experiment we investigate the performance
of the class ORD of order-by queries, and their versions asking for
the first 10 tuples only, see Figure~\ref{fig:exp-sort}. FDB
restructures the factorisation whenever necessary before enumeration
in the required order. The time required for the restructuring is
essentially captured by the execution time of the {\em limit} variant,
since enumerating the first 10 tuples only adds a small constant
overhead.

%%%%%%%%%%%%%%%%%%%%%%%%%%%%%%%%%
%%%%%%%%%%%%%%%%%%%%%%%%%%%%%%%%%

Query $Q_{10}$ asks for a specific order on the materialised view
$R_2$ that is already sorted in that order. FDB thus needs no
restructuring of the view and enumerates the results. The relational
engines need no additional sorting and only scan the relation $R_1$.
This allows us to see how relation scanning compares against FDB
enumeration: The latter is about the same speed as the PSQL scan.
SQLite is faster than both; this is possibly because FDB is not
optimised for string manipulation (by, e.g., hashing all strings in a
relation). Returning the first 10 tuples only takes negligible time
for both FDB and PSQL, SQLite takes longer.

Query $Q_{11}$ asks for a slightly different order than the one
existing in the relational input $R_2$. FDB need not do any work,
since its factorisation of $R_2$ does already support this new order,
as well as the original order.  Simultaneous support for several
orders is a key feature of FDB. The f-tree of the factorised
materialised view $R_1$ is $\T_1$ from the introduction, where we have
package instead of pizza. It can therefore support both orders
(package, date, item) and (package, item, date).  In contrast, both
relational engines need to sort the relation from scratch. Enumeration
with FDB takes the same time as for $Q_{10}$, yet now the relational
engines need more time to sort, PSQL about one order of magnitude
more. Even to return the first 10 tuples requires major work for the
relational engines, while FDB returns each of these tuples with
constant delay and no precomputation.

Query $Q_{12}$ asks for an order that is not already supported by the
f-tree of the factorised materialised view. In this case, FDB needs
restructuring before the enumeration: one swap between date and its
parent node package is enough, and is still faster than sorting from
scratch using either of the relational engines. Returning the first 10
tuples in the required order using PSQL takes the same time as the
swap.

Query $Q_{13}$ just sorts the relation $R_3$. Remarkably, even for
sorting a (non-factorised) relation, FDB outperforms the relational
engines since it only needs to partially re-sort the input. This is
achieved by swapping the attributes date and customer. The
factorisation constructed by FDB groups the relation by the first
attribute in the sorting order, then by the second, and so on.  The
swap of date and customer re-groups by customer instead of date, yet
the list of packages for each date and customer remains sorted.

%%%%%%%%%%%%%%%%%%%%%%%%%%%%%%%%%%%%%%%%
{\noindent\bf Experiment 5: Overhead of relational engines.}  PSQL and
SQLite are full-fledged engines while FDB is not. To understand their
overhead, we also benchmarked a basic main-memory relational engine
called RDB; this has been previously used for benchmarking against
FDB~\cite{BOZ12} for select-project-join queries and we extended it
with sorting and aggregation operators for our purpose. We ran all
queries in the previous experiments also with RDB. Where grouping is
required, RDB first sorts the records (using C++ STL sort) and then
performs aggregation in one scan. We found that RDB's performance is
very close to SQLite's (which implements grouping by sorting using
B-trees) and we therefore not explicitly show it in the plots.

%\nop{
%fdb with f-paths:
%- the data from tim's machine is bad because of other stuff running, I noticed this late and asked giovanni whether we could have the machine (no response yet, well, sunday)
%- the data we have indicate that the times are by a constant factor (3-5?) better than for psql. my explanation is that we already have implicit grouping in the data, so we do not need to do the group-by
%- for Q_2, the aggregate on the f-path package-date-customer-item-price requires to swap customer upwards twice, but this creates branching, because the operator automatically checks for independent subtrees. thus we do not really simulate flat processing only

%this highlights a point which we never really spell out: FDB has some explicit grouping in the data, so the group-by operation is not needed once we have an f-tree compatible with the group-by.
%}

%% file: relatedwork.tex
% !TEX root = main.tex

\section{Related Work}
\label{sec:related}

%%%%%%%%%
\nop{
Taken from the introduction:

This idea is reminiscent of eager aggregate evaluation in the
relational setting~\cite{YanL:95}, which {\em partially} pushes
aggregation past joins.

 \new{To compute this factorisation,
we thus first evaluated the join and then the aggregate. In contrast
to the relational case, pushing the aggregate past
joins~\cite{YanL:95} has less impact on performance since the
factorisation of the join result $R$ is at most linear in the size of
the input even if the join}

A similar optimisation can also be
achieved in the relational setting by pushing the aggregate past the
join, provided $R$ is not already materialised: we aggregate after
joining Pizza and Items and before joining with Orders.
}
%%%%%%%%%

\nop{
TO BE MERGED IN:

If both the input and output are not factorised, then the optimised
evaluation of FDB can be \emph{simulated} using purely relational
processing, \new{such simulation encompasses several relational
optimisation techniques. Most importantly, FDB combines the advantages
of both lazy aggregation (speedup brought by executing a selective
join first) and eager aggregation~\cite{YanL:95} (data reduction by
preaggregation before join) and applies the techniques of partial
preaggregation to an arbitrary number of joins.} For queries with
order-by clauses, our techniques on partial restructuring for changing
the sorting order can benefit the relational setting as well.

}

There is a wealth of related work on storage layout, succinct data
representations, schema design, polynomial-delay enumeration for query
results, and aggregate processing.

{\bf\noindent Storage layout.}  \nop{One active direction of work is
  on columnar stores and horizontal partitioning.}  Similar to FDB,
columnar stores, e.g., MonetDB~\cite{MonetDB99} and
C-Store~\cite{MonetDB99,SAB+2005}, target read-optimised scenarios.
Horizontal partitioning or sharding is used for data distribution and
can increase parallelisation of query processing. Partitioning-based
automated physical database
design~\cite{Agrawal:Partitioning04,Hyrise2010} has been proposed for
maximising the performance of a particular workload. RodentStore is an
adaptive and declarative storage system providing a high-level
interface for describing the physical representation of
data~\cite{RodentStore2009}.  In contrast to existing approaches, FDB
intertwines vertical and horizontal partitioning of relational data.
For this reason, existing techniques are not directly applicable.

{\bf\noindent Data compression.} Data compression shares with
factorisation the goal of compact data representation, as used e.g.,
for column compression~\cite{SAB+2005,Hyrise2010} and
dictiona\-ry-based value compression in
Oracle~\cite{CompressionOracle03}\nop{, and the XMill compressor of
  XML data~\cite{Liefke:CompressedXML:2000}.  Structure-preserving
  compression of trees into directed acyclic graphs has been used to
  boost the performance of XML query
  evaluation~\cite{CK:CompressedXML:2003}}. Such data compression
schemes can benefit FDB and complement the structural compression
brought by factorised representations.

{\bf\noindent Schema design.}  Factorisation trees rely on join
dependencies, which form the basis of the fifth normal
form~\cite{Gehrke2003}. Join dependencies were not used previously as
a basis for a representation system for relational data that can
support query processing. Factorisations can go beyond the class of
factorisation trees and the query processing techniques developed in
this paper can be adapted to more general factorisations.

{\bf\noindent Succinct representation systems and applications.}
Factorised databases have been introduced recently~\cite{OZ12,BOZ12}.
Generalised hierarchical decompositions~\cite{Delobel78} and compacted
relations~\cite{BRS:VLDB:1982} are equivalent to factorisations over
f-trees but questions of succinctness have not been addressed by
earlier work. Nested relations~\cite{Makinouchi1977,JS1982,Verso1986}
are also structurally equivalent to factorisations over f-trees, their
data model is explicitly non-first normal form. Previous work does not
study representing standard relations by nested relations, nor the
related questions of choosing a succinct representation and evaluating
queries on the represented relation.

In provenance and probabilistic databases, factorisations can be used
for compact encoding of provenance
polynomials~\cite{Green:PODS:2007,OZ11b} and for efficient query
evaluation~\cite{PDB-BOOK11}. They can be used to represent large
spaces of possibilities or choices in design
specification~\cite{INV1991} and in incomplete
information~\cite{OKA08gWSD}.

{\bf\noindent Aggregate processing.}
\nop{Our approach to partial
aggregation is related to work by Yan and Larson on equivalences for
queries with aggregates~\cite{YanL:95}. They show how sum and count
aggregates can be partially pushed past a join. A similar technique
was later proposed for a more complex aggregate that computes exact
probabilities in probabilistic databases~\cite{OHK2009}. While these
approaches rely on query rewriting, our approach conveys information
about partial aggregation in the f-trees of temporary results, and
replaces elaborate rewrite rules by simple compositional rules for
aggregation operators.}
Our approach to partial aggregation before restructuring is intimately
related to work by Yan and Larson on partially pushing $sum$ and
$count$ aggregation past joins ~\cite{YanL:95}. This is called eager
aggregation and contrasts with lazy aggregation, which is applied
after joins. While their technique relies on query rewriting, our
approach conveys information about partial aggregation in the f-trees
of temporary results, and replaces elaborate rewrite rules by simple
compositional rules for aggregation operators. In relational
processing, eager aggregation reduces the size of relations
participating in a join and prevents unnecessary computation of
combinations of values that are later aggregated anyway. FDB already
avoids the explicit enumeration of such combinations by means of
factorisation, whose size is at most the size of the join input and
much less for selective joins. FDB thus combines the advantages of
both lazy and eager aggregation.

{\bf\noindent Enumeration of query results.}  Factorised
representations of query results allow for constant-delay enumeration
of tuples. For more succinct representations, e.g., binary join
decompositions~\cite{Gottlob11} or just the pair of the query and the
database~\cite{Durand07}, retrieving any tuple in the query result is
NP-hard. Factorised representations can thus be seen as compilations
of query results that allow for efficient subsequent processing.
There has been no previous work on enumeration in sorted order on
factorised data. The closest in spirit to ours is on polynomial-delay
enumeration in sorted order for results to acyclic conjunctive
queries~\cite{Kimelfeld06}.

%% file: conclusion.tex
% !TEX root = main.tex

\section{Conclusion and Future Work}
\label{sec:conclusion}

In this paper we introduce processing techniques for que\-ries with
aggregates and order-by clauses in factorised data\-bases.  These
techniques include partial aggregation and constant-delay enumeration
for query results and are implemented in the main-memory query engine
called FDB.  We show experimentally that FDB can outperform the
open-source engines SQLite and PostgreSQL if its input is represented
by views materialised as factorisations.

An intriguing research direction is to go beyond factorisations
defined by f-trees and consider more succinct representations such as
decision diagrams\nop{ or those inspired by rectangle coverings for
algebraic factorisation of logic functions~\cite{brayton87}}.

\nop{
stopping after the first $k$ tuples in the enumeration phase. More
general methods utilising enumeration with larger than constant delay
while needing less restructuring, thus saving time for small $k$, are
subject to future work.
}